\documentclass[citesort,number,dvips]{arxstspdf}
\usepackage[vtexbibl]{}
\usepackage{stfloats}
\usepackage{graphicx}
\usepackage{flushend}

\volume{24}
\issue{4}
\pubyear{2009}
\firstpage{517}
\lastpage{529}
\doi{10.1214/09-STS306}



\begin{document}
\begin{frontmatter}

\title{Estimating Effects and Making Predictions from Genome-Wide
Marker Data}
\runtitle{Predictions from Genome-Wide Marker Data}

\begin{aug}
\author[a]{\fnms{Michael E.} \snm{Goddard}},
\author[b]{\fnms{Naomi R.} \snm{Wray}},
\author[c]{\fnms{Klara} \snm{Verbyla}}
\and
\author[d]{\fnms{Peter M.} \snm{Visscher}\ead[label=e4]{Peter.Visscher@qimr.edu.au}\corref{}}
\runauthor{Goddard, Wray, Verbyla and Visscher}

\affiliation{University of Melbourne, Queensland Institute of Medical Research, University of Melbourne
and Queensland Institute of Medical Research}

\address[a]{Michael E. Goddard is Professor of Animal Genetic,
Faculty of Land and Food Resources, University of Melbourne and
Department of Primary Industries, Victoria, Australia.}
\address[b]{Naomi R. Wray is Professor of Psychiatric,
Genetic Epidemiology and Queensland Statistical Genetics,
Queensland Institute of Medical Research, Australia.}
\address[c]{Klara Verbyla is PhD Scholar,
Faculty of Land and Food Resources, University of Melbourne and
Department of Primary Industries, Victoria, Australia.}
\address[d]{Peter M. Visscher is Professor of Statistical Genetics,
Genetic Epidemiology and
Queensland Statistical Genetics, Queensland Institute of Medical
Research, Australia \printead{e4}.}

\end{aug}

\begin{abstract}
In genome-wide association studies (GWAS), hundreds of thousands of
genetic markers (SNPs) are tested for association with a trait or
phenotype. Reported effects tend to be larger in magnitude than the true
effects of these markers, the so-called ``winner's curse.'' We argue
that the classical definition of unbiasedness is not useful in this
context and propose to use a different definition of unbiasedness that
is a property of the estimator we advocate. We suggest an integrated
approach to the estimation of the SNP effects and to the prediction of
trait values, treating SNP effects as random instead of fixed effects.
Statistical methods traditionally used in the prediction of trait values
in the genetics of livestock, which predates the availability of SNP
data, can be applied to analysis of GWAS, giving better estimates of the
SNP effects and predictions of phenotypic and genetic values in
individuals.
\end{abstract}

\begin{keyword}
\kwd{Genome-wide association study}
\kwd{prediction}
\kwd{estimation}.
\end{keyword}

\end{frontmatter}

\section{Introduction} \label{sec:1}

The rules for the genetic inheritance of traits, discovered by Mendel,
are most obvious for traits controlled by a single gene, for example,
individuals who carry two defective variants in the gene CFTR develop
cystic fibrosis. However, most of the traits that are of importance in
medicine, agriculture and evolution are influenced by many genes and by
nongenetic or ``environmental'' factors. For example,\break height in humans
involves many physiological processes and many genes but is also
influenced by nongenetic factors such as nutrition and health care.
These traits are called quantitative or complex traits and include
common genetic diseases such as heart disease, breast cancer, diabetes
and psychiatric disorders.

Until recently few of the genes which harbor variants for complex
genetic traits had been identified. The availability of genome-wide
panels of densely spaced, genetic markers has led to a revolution in the
study of the genetics of complex traits. These genetic markers are
single nucleotide polymorphisms (SNPs) which are positions in the DNA
sequence where the nucleotides can vary (e.g., G or T). Individuals
carry pairs of homologous chromosomes and so have one of three genotypes
at a G/T SNP---GG, GT or TT. Assays are now available that determine
the genotype of an individual at 100,000 to over 1 million SNPs spread
over all of the chromosomes of the species.

SNPs usually have no direct effect on a trait under study. However, any
polymorphism that does affect the trait will be located on a chromosome
close to one or more of the genotyped SNPs because the genotyped SNPs
are chosen to cover all chromosomes in, at least, moderate density.
Polymorphisms that are located close to each other on a chromosome can
occur together more often than expected by chance, so that they are
correlated or in linkage disequilibrium (LD). Thus, for every
polymorphism that affects a trait, there is likely to be a SNP nearby
that is in LD with the causal polymorphism and hence correlated or
associated with the trait. Since the SNPs cover the whole genome,
experiments that test for association between a trait and a panel of
SNPs that cover the whole genome are called genome wide association
studies (GWAS). GWAS have discovered numerous SNPs that are associated
with, for example, complex diseases such as Crohn's disease and type II
diabetes \cite{1,2}. Typically there are multiple SNPs associated with a
complex trait, each one with a small effect, or a small increase in the
risk of a particular disease.

One purpose of a GWAS might be to find the genes and polymorphisms that
affect the trait. Hopefully this will elucidate the biology of the trait
and, in human medicine, may lead to new therapeutics. In this case we
would like to have unbiased and accurate estimates of the effects of
SNPs on the trait on which to base further experimentation. Another use
of the GWAS is to use the SNPs to predict the phenotypic or genetic
value of individuals. For example, in agriculture it would be very
useful to predict the genetic merit of bulls for milk production using
DNA markers such as SNPs, because it is not possible to observe the
phenotype (milk yield) in bulls and, even if it were possible, it is the
genetic value of the bull that will be passed on to his descendants.
Also, if we could predict the risk of a specific disease in individuals
based on DNA markers this would be useful in diagnosis, treatment,
prognosis and prevention. The DNA markers cannot predict the
environmental effect on a complex trait, only the genetic value for that
trait. Hence, the genetic value and phenotypic value predicted from DNA
markers are the same. The question to be considered is as follows: how
can this prediction be made as accurately as possible?

Statistical analysis of GWAS may test hypotheses (e.g., there are no
SNPs associated with this trait) or estimate the effect of a SNP on the
trait (e.g., how much does this SNP affect the probability that a person
will develop diabetes). In such estimation problems the tendency has
been to treat the effect of a SNP as a fixed effect and use estimators
that are classically unbiased, at least approximately, such as the
maximum likelihood estimate of the relative risk of disease. However, in
GWAS hundreds of thousands of SNPs are tested, but frequently the
estimated effects are reported only for the significant markers, for
example, \cite{1}. Under these conditions, it has been known for some time
that the reported effects tend to be larger in magnitude than the true
effects of these markers. This effect is known as the ``Beavis effect'' in
agricultural genetics \cite{3,4} as cited in \cite{5}, and has been described as
a form of the ``winner's curse'' \cite{6}. Methods to correct for this bias
have been published \cite{3,6,7,8,9,10,11,12,13,14,15}. We argue that the classical definition of
unbiasedness is not useful in this context and propose to use a
different definition of unbiasedness that is a property of the estimator
we advocate in this paper. This definition of unbiasedness has a strong
theoretical underpinning \cite{16,17} and has traditionally been used and
applied in agricultural genetics \cite{16,17,18,19,20}.

Another motivation for the statistical analysis of GWAS might be to use
the SNP genotypes to predict the value of a trait that has not yet been
observed, for example, to predict the future risk of a disease for an
individual person. This has parallels to predicting a person's risk of
disease based on their family history of that disease. Generally, when
such prediction is carried out, the variable being predicted is regarded
as a random variable. Then the prediction might use the multiple
regression of the trait or phenotypic value on the SNP genotypes. If the
biased estimates of the SNP effects described in the previous paragraph
are used in this regression equation, then the predictions will
exaggerate the variation in risk between individuals.

In this paper we suggest an integrated approach to the estimation of the
SNP effects and to the prediction of trait values that overcomes the
bias in both. It relies on treating the SNP effects as random, instead
of fixed, effects. There is a well-established statistical tradition of
prediction of trait values in the genetics of livestock and we introduce
this methodology in the first section of the paper. This methodology
predates the availability of SNP data and uses the equivalent of family
history, that is, phenotypic values on relatives of the individual whose
phenotype we wish to predict. Then we show how this approach can be
applied to analysis of GWAS giving better estimates of the SNP effects
and predictions of phenotypic and genetic values in individuals. In
fact, there is equivalence between models of genetic value based on SNPs
and one based on relationships between individuals, and this equivalence
is explained. We then give results from the analysis of GWAS. Finally we
discuss how this approach will cope with future developments such as
whole genome resequencing.

\section{Prediction} \label{sec:2}

\subsection{Prediction of Phenotypic Values from Pedigree Data} \label{sec:2.1}

To illustrate our approach, we will describe a specific example and then
generalize. Imagine that we have data consisting of the milk yields of
cows that belong to a number of half-sib families (all cows within a
half sib family have the same sire) and we wish to predict the milk
yields of future cows within any of these half-sib families. The milk
yields of cows within a family are correlated and we can use this to
predict the yield of a future cow from the same family
($y_{\mathrm{future}})$. In general, the predictor of a random variable that
has the lowest mean square error of prediction is the conditional mean
given the data available for prediction \cite{16,17,19,21}. This predictor
is called the best predictor \cite{17,20}. In the case of predicting the
milk yield of a future cow conditional on the milk yields observed on
existing cows in the same family ($\mathbf{y}_{\mathrm{existing}}$), the best
predictor ($\hat{y}$) is the expected value of the milk yield of a
future cow. That is,
\[
\hat{y}= \mathrm{E}(y_{\mathrm{future}} |\mathbf{y}_{\mathrm{existing}}).
\]
If the $y$ follows a multivariate normal distribution, then
E($y_{\mathrm{future}} |\mathbf{y}_{\mathrm{existing}})$ is the linear regression
of\break $y_{\mathrm{future}}$ on $\mathbf{y}_{\mathrm{existing}}$. If
\begin{eqnarray*}
V(\matrix{ y_{\mathrm{future}}& \mathbf{y}_{\mathrm{existing}}}) = \mathbf{V} =\left[
\matrix{ {V}_{\mathrm{future}} & \mathbf{v}
\cr
\mathbf{v}^{\prime} & \mathbf{V}_{\mathrm{existing}}
}
\right],
\end{eqnarray*}
then $\hat{y}=\mathbf{v}^{\prime}\mathbf{V}_{\mathrm{existing}}^{ -
1}y_{\mathrm{existing}}$. Since all cows within a family share
the same relationship with each other, the diagonal elements of
$\mathbf{V}$ are all equal, as are the off-diagonal elements. That is,
$\mathbf{V} = \mathbf{I}\sigma_{e}^{2} + \mathbf{J}\sigma_{s}^{2}$,
where $\mathbf{I}$ is the identity matrix, $\mathbf{J}$ is a matrix of all
ones, $\sigma_{e}^{2} + \sigma_{s}^{2}$ is the variance of milk yield
and $\sigma_{s}^{2}$ is the covariance between the milk yields of cows
from the same family.

An equivalent model, generalized to $n$ milk yield records from $f$ sire
families, that leads to the same prediction is as follows:
\[
\mathbf{y} = \mathbf{Zs} + \mathbf{e},
\]
where
\begin{itemize}
\item[]$\mathbf{y}$ is an $n \times  1$ vector of milk yields for all cows over $f$
families,

\item[] $\mathbf{Z}$ is an $n \times  f$ matrix that allocates cows to
families,

\item[] $\mathbf{s}$ is an $f \times 1$ vector of sire effects $\sim \mathrm{N}(0,
 \mathbf{I}\sigma_{s}^{2})$,

\item[] $\mathbf{e}$ is an $n \times 1$ vector of independent errors or environmental
effects $\sim\mathrm{N}(0, \mathbf{I}\sigma_{e}^{2})$.
\end{itemize}
The best predictor of $\mathbf{s}$ is
\[
\hat{\mathbf{s}}= \mathbf{Z}'(\mathbf{ZZ}' \sigma_{s}^{2} +
\mathbf{I}\sigma_{e}^{2})^{-1} \mathbf{y}
\]
and the best predictor of $y$ for a future cow is
\[
\hat{y}= \mathbf{z}_{\mathrm{future}}^{\prime}\hat{\mathbf{s}},
\]
where $\mathbf{z}_{\mathrm{future}}$ is a vector of zeros with a single one, to
indicate to which family the future cow belongs.

For the $i${th} sire, $\hat{s}_{i}$ can also be written as $y_{i}
n_{i} /(n_{i}+ \lambda))$, where $y_{i}$ is the mean of $y$ for the
$n_{i}$ cows in the $i${th} family and $\lambda =
\sigma_{e}^{2}/\sigma_{s}^{2}$. This is a linear model and $\hat{s}_{i}$ is an
estimate of the sire effect ($s_{i}$), but it is not the conventional
estimate derived from treating sire effects as fixed effects which would
be $\tilde{s}_i= y_{i}$. Estimates such as $\tilde{s}$ are unbiased in the
traditional sense, that is, they have the property
\begin{equation}\label{eq1}
\mathrm{E}(\tilde{s}|s) = s.
\end{equation}
By contrast, $\hat{s}$ is not unbiased in the sense of (\ref{eq1}) because it is
a ``shrunk'' estimate.

In what respect is $\hat{y}$ the best predictor of $y?$ It has the
minimum prediction error variance $\operatorname{var}(\hat{y}- y)$. Similarly,
$\hat{s}$ has the minimum $\operatorname{var}(\hat{s}- s)$. $\hat{s}$ also has the
property \cite{16,19,20,21}
\begin{equation}\label{eq2}
\mathrm{E}(s |\hat{s}) =\hat{s}.
\end{equation}
Equation (\ref{eq2}) defines a type of unbiasedness which is only meaningful
when $s$ is regarded as a random effect. It can be stated as follows:
If, on the basis of this analysis, one selects a group of sires whose
average value of $\hat{s}$ is $k$ and one produces one more daughter from
each of these sires, then the expected mean milk yield of these future
daughters is $k$.

In practice, the statistical model for milk yields would include some
fixed effects ($\mathbf{c}$), as well as the random effect of sire, in a
mixed model. That is,
\[
\mathbf{y} = \mathbf{Xc} + \mathbf{Zs} + \mathbf{e}.
\]
Also, the sires might be related and so $\mathbf{s} \sim \mathrm{N}(0,
\mathrm{A}_{s}\sigma_{s}^{2})$, where $\mathbf{A}_{s}$ is the numerator
relationship matrix (which is twice the kinship matrix \cite{18}) derived
from the pedigree of the animals. The solutions from the equations
\[
\left[ \matrix{ \mathbf{X'X} & \mathbf{X'Z} \cr \mathbf{Z'X} &
\mathbf{Z'Z + A}_{{s}}^{ - 1}\lambda
} \right]^{ - 1}\left[
\matrix{ \mathbf{X'y} \cr \mathbf{Z'y} } \right]
\]
are Best Linear Unbiased Estimates (BLUEs) of the fixed effects and Best
Linear Unbiased Predictors (BLUPs) of the random effects \cite{16,20}. In
fact, we can model the observed phenotypes in terms of the genetic value
of each individual ($\mathbf{a}$) as
\[
\mathbf{y} = \mathbf{Xc} + \mathbf{Ta} + \mathbf{e},
\]
where $\mathbf{a}$ is a vector of additive genetic values\break $\sim$$ \mathrm{N}(0,
\mathbf{A}\sigma_{{a}}^{2})$ where $\mathbf{A}$ is a numerator
relationship matrix like $\mathbf{A}_{s}$ but recording the
relationships between individuals, including the relationships caused by
cows with the same sire. The BLUP equations become
\[
\left[ \matrix{ \mathbf{X'T} & \mathbf{X'T} \cr \mathbf{T'X} &
\mathbf{T'T} + \mathbf{A}^{ - 1}\frac{\sigma _{e}^{2}}{\sigma _{a}^{2}}
} \right]^{ - 1}\left[ \matrix{ \mathbf{X'y} \cr \mathbf{T'y}
} \right].
\]
In livestock genetics this is known as an ``animal model'' because the
genetic value of each individual is explicitly included in the model
together with the relationship between all animals. This model has also
been used in evolutionary genetic studies \cite{22} and in QTL linkage
mapping studies in human pedigrees \cite{23}. For linear mixed models
containing fixed and random effects, property (\ref{eq2}) holds when the
estimates are obtained using BLUP, provided the effects are normally
distributed with known variances \cite{16,17}.

\subsection{Prediction of Phenotype from SNP Genotypes and
Estimation of SNP Effects} \label{sec:2.2}

In the analysis of a GWAS we would like to both predict the phenotype of
individuals with observed genotypes but no observed phenotype and to
estimate the effects of the SNPs. Both these objectives can be achieved
in a consistent manner if we treat the SNP effects as random, just as we
did the sire effects above. The structure of the data is typically as
follows. There is a reference or discovery sample of individuals who
have been typed for the SNPs and recorded for the phenotype. From this
data a prediction equation is derived that predicts phenotypes from SNP
genotypes. This prediction equation is then used on a validation sample
and the accuracy of prediction that we wish to maximize is the accuracy
of predicting phenotypes in this sample.

What properties would an ideal predictor have and what measure of
accuracy should be maximized? We propose that the ideal predictor is the
expectation of the phenotype conditional on the SNP genotypes. That is,
if the phenotype is $y$, the predictor $\hat{y}$ should be
\[
\hat{y}= E(y |\mathrm{SNP} \mbox{ genotypes}).
\]
This property has many advantages. It maximizes the correlation between
$y$ and $\hat{y}$, minimizes the error mean square and results in a
regression of $y$ on $\hat{y}, \beta(y,\hat{y}) = 1$ \cite{16,20}.
Consistent with this approach, we would estimate the SNP effects ($b$)
such that
\[
\hat{b}= E(b |\mbox{ data on phenotypes and genotypes}).
\]
This estimator is also unbiased in the sense of (\ref{eq2}) and has the same
properties as listed for $\hat{y}$ above.

\subsection{Comparison with Traditional Fixed Effect Estimators} \label{sec:2.3}

\begin{figure*}[t]

\includegraphics{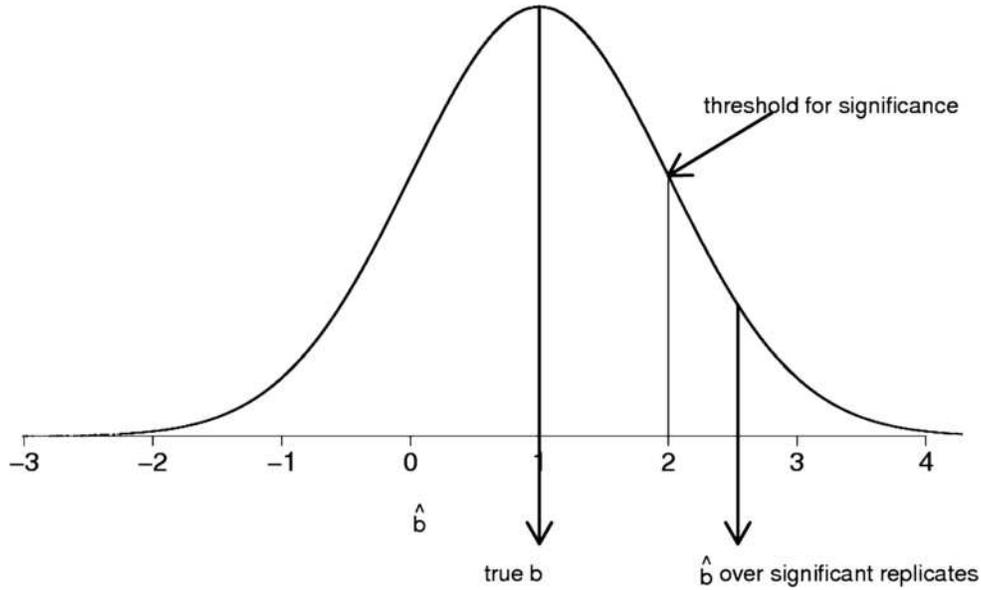}

  \caption{The mean value of $\hat{b}$ ($\sim 2.5$) from significant
replicates when $b = 1.0$, the threshold for significance is 2.0 and the
SE of $\hat{b}$ is 1.0.}\label{fig1}\vspace*{6pt}
\end{figure*}

Usually the effects of SNPs in GWAS have been estimated by methods that
are unbiased in the traditional sense of having property (\ref{eq1}). To
demonstrate that an estimator is unbiased in this sense, we would test
the same marker in numerous replicate experiments and average the
estimates over these replicates. The simple least squares estimate of
$b$ is unbiased in this sense. (We use ``least squares'' throughout as an
example of an estimation procedure that does not shrink estimates. Other
estimation procedures, for example, maximum likelihood to estimate odds
ratios using logistic regression, fall in the same category.) However,
this unbiased property is lost if we only average the estimates over the
replicates in which the marker effect was declared ``significant''
according to some arbitrary threshold of the test statistic. This effect
is illustrated in Figure~\ref{fig1} where a SNP with effect 1.0 and standard
error 1.0 is declared significant only if the estimated effect $\hat{b}$
exceeds~2.0. In those significant replicates, the mean estimate of $b$ is
$\sim$2.5. Methods such as \cite{6,8,24} attempt to correct for this bias so
that the average of $\hat{b}$ over the \textit{significant} replicates is $b$.

\begin{figure*}[b]

\includegraphics{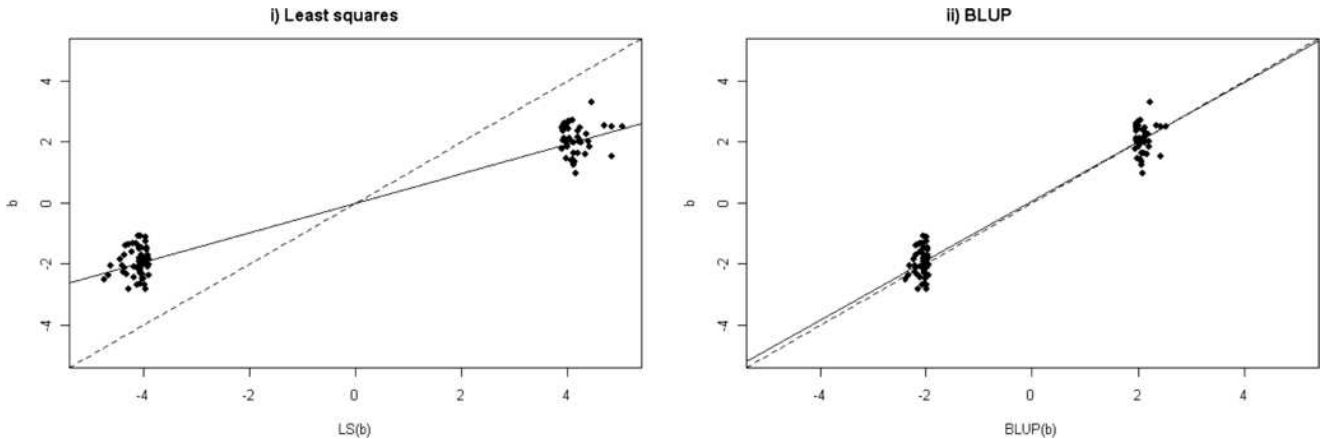}

  \caption{Comparison of conventional least squares (\textup{LS}) estimates \textup{(i)} and
BLUP estimates \textup{(ii)} of the effects of SNPs. The true SNP effects were
simulated $\sim$\textup{N}(0, 0.5) and estimated with sampling $\mathrm{error} \sim\mathrm{N}(0,
0.5)$. The SNPs with the largest magnitude of effect are plotted
($|\operatorname{LS}(b)| > 4$). The BLUP estimates are unbiased, while the least
squares estimates overestimated the magnitude of the largest effects.
The dashed line shows $y = x$ and the solid line is the regression of $b$
on the $\operatorname{LS}(b)$ or $\operatorname{BLUP}(b)$. BLUP estimates are unbiased irrespective of
the threshold chosen for selection.}\label{fig2}
\end{figure*}

Conversely, we recommend estimators of SNP effects which are unbiased in
the sense of (\ref{eq2}) rather than the usual estimators which are unbiased in
the sense of (\ref{eq1}). Which type of unbiased estimator do we want? We argue
that unbiasedness of type (\ref{eq2}) is more useful. Estimators of type (\ref{eq2})
maximize the correlation between $b$ and $\hat{b}$, minimize the error
mean square and result in a regression of $b$ on $\hat{b},
\beta(b,\hat{b}) = 1$. Estimators of type (\ref{eq1}) do not have these
properties because $\operatorname{var}(\hat{b}) > \operatorname{Cov}(b,\hat{b})$, that is, the
variance of the predictor is larger than its covariance with the true
value.

Generally a GWAS, using 100,000s of markers, is not conducted to
estimate the effect of a single marker. If it was and we wanted an
unbiased estimate of its effect in the sense of (\ref{eq1}), we could simply use
the least squares estimate regardless of whether or not it was
significant. Generally, the GWAS is used to select markers which have
the largest or most significant effects (we will assume $b > 0$ since
the sign is arbitrary). The markers might be selected for further
experimentation or for use to predict disease risk \cite{25}. When estimators
of type (\ref{eq2}) are used to select the ``best'' markers, this maximizes the
mean true effect of the group of selected markers \cite{26}. In addition, the
expected mean true value of $b$ in this group of markers is equal to the
mean $\hat{b}$ of the selected markers. Hence, $\mathrm{E}(b| \hat{b} ) =
\hat{b}$. This is illustrated by simulation in Figure \ref{fig2}. Here 100,000
markers effect ($b$) and their least squares estimates ($\tilde{b}$) are
simulated by $\tilde{b} = b + e$ , $b \sim
\mathrm{N}(0,\sigma_{{b}}^{2})$, $e \sim \mathrm{N}(0,\sigma_{{e}}^{2})$.
We arbitrarily chose $\sigma_{{b}}^{2} = \sigma_{{e}}^{2}
= 1/2$. $\tilde{b}$ is an unbiased estimator of $b$ in the
classical sense that $\mathrm{E}(\tilde{b}| b) = b$. However, if we now select
the SNPs with the largest $|\tilde{b}|$ $(|\tilde{b}| > 4)$, then this
over-estimates the true value $|b|$. This is an example of the winner's
curse and occurs because $\tilde{b}$ is not unbiased in the sense of
(\ref{eq2}). On the other hand, if we estimate $b$ by $\tilde{b} =
\tilde{b}/(1+\lambda)$, with $\lambda =\frac{\sigma _{e}^{2}}{\sigma
_{b}^{2}}$, we see that the average value of $|\tilde{b}|$ among the
selected SNPs is equal to the true average value of $|b|$ because
$\tilde{b}$ is unbiased in the sense of (\ref{eq2}). That is, estimators of the
kind recommended here do not suffer from the winner's curse. This
property holds irrespective of the threshold chosen to select the SNPs.

The advantages of the properties of type (\ref{eq2}) estimators [i.e., those
with property (\ref{eq2})] can be illustrated using two examples. First, suppose
the purpose of selecting the markers is to predict the disease risk
faced by individual people. It is known from prediction theory that when
estimators of type (\ref{eq2}) are used to predict the phenotype of individuals,
this maximizes the correlation between predicted risk and true risk
\cite{26}. Second, suppose we want to design a validation experiment to
confirm the effects of the selected markers and must decide on the size
of this experiment. The design needed for a given power depends on the
true effect of the selected markers. If conventional estimates such as
$\tilde{b}$ are used, we will overestimate the size of effect and so
design an experiment which is too small to detect the true effects.
However, an estimate with property (\ref{eq2}) does not overestimate the
magnitude of the effects and so will lead to a design appropriate to
detect the true effects. In fact, in general, if some decision is to be
made (such as to invest time and money in further experiments) and if
there is a cost to making a bad decision, then estimators with property
(\ref{eq2}) are desirable \cite{27}. For instance, estimators with property (\ref{eq2}) are
used extensively in artificial selection programs in agriculture to rank
individuals on genetic merit \cite{17,28} because this maximizes the genetic
improvement when the highest ranking animals are used as parents of the
next generation \cite{26}.

Thus, property (\ref{eq2}) is useful and it is also, we believe, what scientists
implicitly mean when they use the word ``unbiased'' in this context. In
classical or frequentist statistics we usually seek estimators of fixed
effects with property (\ref{eq1}) and predictors of random effects with property
(\ref{eq2}). For example, BLUP is a well-established method for simultaneous
estimation fixed effects and prediction of random effects \cite{18,19}.
Bayesian statisticians treat all effects as random and so usually seek
estimators with property (\ref{eq2}) for all effects. Regardless of whether one
is a frequentist or Bayesian, if one is going to select the best markers
from among many tested and use them for some purpose, then it is best to
use an estimator with property (\ref{eq2}) for the reasons discussed above.

Another advantage of estimators derived from (\ref{eq2}) is that they can be
applied to the joint data obtained from the initial GWAS and the
validation study and yield estimates which are still unbiased of type
(\ref{eq2}). The estimator $\hat{b}= \mathrm{E}(b |$  data on phenotypes
and
genotypes)
is unbiased in the sense of (\ref{eq2}) by definition, regardless of the amount
of data used. That is, one does not need to distinguish between the
discovery data and the validation data. Therefore, in a two stage
experiment in which the markers with largest estimated effect are chosen
from stage 1 and further data collected on them in stage 2, all the data
can be combined into one data set and analyzed without discriminating
the stage 1 and stage 2 data. In contrast, classical estimates of SNP
effects combining the initial GWAS and the validation study (as in \cite{29})
will still be biased for SNPs selected on the basis of the GWAS.

\section{Models for Prediction of SNP Effects} \label{sec:3}

Three difficulties underlie the prediction of SNP effects and
phenotypes. First, the number of SNPs ($p$) is typically 10--100 times
the number of individuals ($n$) in the sample, the so called $p > n$ (or
large $p$ small $n$) problem, which leads to difficulty of an
oversaturated model. Second, it is not the SNPs that are genotyped that
directly affect phenotype but unknown polymorphisms that are often
called quantitative trait loci or QTL. Third, the QTL may affect
phenotype in a complicated and unknown manner. For example, the QTL may
interact in their effects on phenotype (a phenomenon called epistasis).

\subsection{Linear Models} \label{sec:3.1}

We will begin by considering only predictors that are linear in the SNP
genotypes.

Let $y = \mathbf{b}'\mathbf{x} + {e},$
where
\begin{itemize}
\item[] $\mathbf{x}$  is a $p \times 1$ vector of SNP genotypes coded 0, 1 or 2
according to the number of copies of an arbitrarily chosen reference
allele,

\item[] $\mathbf{b}$  is a $p \times 1$ vector of regression coefficients. We will call
$b$ the effects of the SNPs on the trait, although in reality it is the
unknown QTL in LD with the SNPs that actually affect the trait, and

\item[] $e$ is an independent error.
\end{itemize}

Then $\hat{y}=\mathbf{\hat{b}}'\textbf{x} = \mathrm{E}(\mathbf{y} |\mathbf{x})$ implies
\begin{eqnarray}\label{eq3}
\mathbf{\hat{b}}= \mathrm{E}(\mathbf{b} |\mathbf{x, y}) &=& \int\mathbf{b}
p(\mathbf{b}) p(y| \mathbf{b},\mathbf{x}) \,d\mathbf{b}
\\
&&{}\Big/ \int
p(\mathbf{b}) p(y| \mathbf{b},\mathbf{x})\, d\mathbf{b},\nonumber
\end{eqnarray}
where
\begin{eqnarray*}
p(\mathbf{b}) &=& \mbox{ the probability density of } \mathbf{b},
\\
p(y| \mathbf{b}, \mathbf{x})&=& \mbox{ the likelihood}.
\end{eqnarray*}
This makes it clear that prediction of $y$ depends on $p(\mathbf{b})$,
that is, the distribution of the ``effects'' of the SNPs. This can be
considered in a Bayesian framework as the prior information or
distribution, but it can also be put in a frequentist framework if the
effects of SNPs are considered a random variable. Since there are
(hundreds of) thousands of SNPs, it is not unreasonable to consider the
effect of any one SNP as being drawn from a distribution of SNP effects
$p(\mathbf{b})$.

Meuwissen et al. (2001) \cite{30} considered several possible forms of
$p(\mathbf{b})$ in a Bayesian framework so that after specifying the
distribution of $\mathbf{b}$, the distribution parameters are estimated
from the data simultaneous with the estimate of individual SNP effects.
If $b \sim \mathrm{N}(0,\sigma_{b}^{2})$, then $\hat{b}$ is the best linear
unbiased predictor or BLUP of $b$ and $\hat{y}$ is BLUP of $y$. This
implies that all SNP effects are drawn from the same distribution. The
total genetic variance explained by the SNPs is $\sigma_{g}^{2} =
\sigma_{b}^{2} \sum 2p_{i}(1-p_{i})$, where $p_{i} = \mathrm{allele}$
frequency at SNP $i$ and summation is across all SNPs. This implies that
many SNPs have a small effect on the trait but none have a big effect.

Alternatively, one can assume $b_{i} \sim \mathrm{N}(0,\sigma_{b_{i}}^{2})$,
where $\sigma_{b_{i}}^{2}$ is drawn from an inverse scaled $\chi^{2}$
distribution. The use of this hyper-prior distribution implies the
assumption that a large number of SNP effects are small or extremely
close to zero with a few larger effects. This assumption is reflected in
the results of Hayes and Goddard (2001) \cite{30} and Weller et al. (2005)
\cite{31} who examined the distribution of QTL effects in livestock
populations. This method (called Bayes A by the authors) can be
implemented using a Gibbs chain. The use of the normal---inverse scaled
$\chi^{2}$ mixture distribution results in a Student $t$ distribution
for $b$. This allows $b$ to have a distribution with a longer tail than
a normal distribution. A third method (called Bayes B by the authors)
described by Meuwissen et al. (2001) \cite{30} had the same assumptions as
Bayes A about a proportion $q$ of the SNPs but, in addition, assumed
$1-q$ of the SNPs have zero effect, such that
\begin{eqnarray*}
\sigma_{b_{i}}^ 2 &\sim&\chi^{ - 2}(r,s)\quad  \mbox{with probability } q,
\\
\sigma_{b_{i}}^{2}&=& 0 \quad  \mbox{with probability } 1-q.
\end{eqnarray*}
The use of this prior means that the dimensionality of the model is
changing as the number of SNPs included in the model varies. As such, a
reversible jump MCMC algorithm \cite{32} is needed to communicate across all
possible models and their differing dimensionality according to the
proper acceptance ratio. This acceptance ratio is identical to that of
the Metropolis--Hastings algorithm when the Jacobian (which appears due
to the deterministic transformation used in the proposal mechanism) is
equal to one. This occurs even though the dimensions are varying because
the Jacobian itself is not an inherent component of the dimension
changing MCMC.

All these methods ``shrink'' the estimate in some way. BLUP is a linear
function of the data and shrinks all estimates with the same standard
error by the same amount, whereas the other methods are nonlinear
functions of the data and shrink small estimates more than big ones.

Other methods such as LASSO (least absolute\break shrinkage and selection
operator) also give shrunk estimates and can sometimes be interpreted as
approximations to (\ref{eq2}). For example, the LASSO approximates the estimates
when the distribution $p(b)$ is a mixture distribution in the form of a
normal exponential resulting in a Laplace (double exponential)
distribution \cite{33,34,35}. Hoggart et al. \cite{36} use a penalized maximum
likelihood approach combined with stochastic search methods to
demonstrate efficient simultaneous analysis of genome-wide SNPs. They
showed that a normal-exponential-gamma prior led to improved SNP
selection in comparison with single-SNP tests. Lewinger et al. \cite{37}
proposed a hierarchical Bayes marker association prioritization to
select markers for subsequent investigation. They used a prior for the
true noncentrality parameter of association with a large mass at zero
and a continuous distribution of values that are nonzero. In simulated
data, methods without an explicit assumption about $p(b)$ have also
performed well. For example, Wray et al. \cite{25} used multiple regression
on only highly significant SNPs and Lande and Thompson \cite{11} used
multiple regression and cross-validation.

\subsection{Nonlinear Models} \label{sec:3.2}

Models that are nonlinear in the SNP effects might be used for two
reasons. First, a combination of SNPs might be a better predictor of the
allele at the QTL than a linear combination of SNPs. The alleles that
occur at adjacent loci on the same chromosome are known as a haplotype.
Considering a group of $m$ SNPs each with two alleles, there are a
maximum of $2^{m}$ haplotypes, although often the number actually
observed is less than this due to LD. For each haplotype, the frequency
of the positive allele at the QTL may vary. In simulated data, Goddard
(1991) \cite{38} showed that haplotypes of markers predicted the QTL allele
better than a linear combination of the markers and in real cattle data,
Hayes et al. (2006) \cite{39} showed that a haplotype predicted the allele at
an additional marker better than the individual SNPs. The value of
haplotypes depends on how well the genotyped SNPs tag the genomic
variation; the use of haplotypes may increase the chance of a tested
variant having the same frequency and being coupled with the causal
variant. However, fitting haplotypes in the prediction equation is
equivalent to fitting the main effects and all interactions among the
SNP alleles on one chromosome and so can reflect a more complex genetic
model. But, fitting haplotypes is not equivalent to fitting all
interactions among genotypes. For example, an individual with the
genotype AT at one SNP and CG at the next could have haplotypes (A-C)
and (T-G) or haplotypes (A-G) and (T-C). An analysis based on genotypes
would not distinguish between these two situations, but one based on
haplotypes would. Although fitting haplotypes implies fitting
interactions between alleles at different SNPs, it is only SNPs close
together on the chromosome that are assumed to interact. Thus, the
haplotype model is a limited nonlinear model, based on the known
biology. The value of the fitting haplotypes is likely to depend on the
structure of the genotyped sample since the use of haplotypes serves to
exacerbate the ``large $p$ small $n$'' problem and unless full pedigrees are
genotyped, haplotypes cannot be estimated without uncertainty.

A second reason for using nonlinear models is that the QTL may interact
in their effects on the trait, and this would generate interactions
among the SNPs in their apparent effects. Such interactions might occur
between QTL or SNPs located anywhere in the genome, so all possible
interactions need to be considered. If there are $10^{6}$ SNPs, there
are $10^{12}$ two-locus interactions and larger numbers of higher order
interactions. Estimating so many effects could decrease the accuracy of
the prediction equation, especially if they are not needed. Interactions
between QTL (epistasis) are known to occur, but the proportion of the
genetic variance due to nonadditive gene action is controversial. Hill
et al. (2008) \cite{40} argue that the nonadditive variance is typically
smaller than the additive genetic variance and, if so, this would
suggest that additive models would be at least the first step in
predicting phenotype. Lee et al. (2008) \cite{41} found no improvement in
their prediction of unobserved phenotypes from genotype data when
fitting epistasis in their models.

As well as interactions between QTL (epistasis), there can be
interactions between alleles at the same QTL (dominance). For example,
if a SNP has alleles T and A, there are three genotypes---AA, TA and
TT. If the mean of the TA individuals is half way inbetween the mean of
the TT and AA individuals for some trait, then the alleles act
additively and there is no dominance. Departure from the additive model
due to dominance can be included but, if the dominance variance is
small, estimation of the additional effects may make the accuracy of
prediction worse instead of better. Lee et al. (2008) \cite{41} found that
including dominance did improve the accuracy of predicting coat color in
mice from SNPs, but this may not be a typical trait because there were a
few genes of major effect segregating in the population. The improvement
in the prediction by fitting dominance for two other quantitative traits
was much smaller.

An alternative form of the nonlinear model is the semi-parametric model
used by Gianola et al. (2006) \cite{42}. They used a reducing Hilbert space
kernel regression and obtained good accuracy of prediction in simulated
data, but they assumed relatively few QTL in their simulation.

\subsection{An Equivalent Model} \label{sec:3.3}

The simple linear model of $p$ SNP effects on $n$ phenotypes
$(\mathbf{y})$, used above, can be written
\[
\mathbf{y} = \mathbf{Wb} + \mathbf{e}
\]
or, equivalently, as
\begin{eqnarray*}
\mathbf{y} &=& \mathbf{a} + \mathbf{e} \quad  \mbox{and}
\\
\mathbf{a} &=& \mathbf{Wb} \quad\mbox{so
that } V(\mathbf{a}) = \mathbf{WW}'\sigma_{b}^{2} = \mathbf{G},
\end{eqnarray*}
where
\begin{itemize}
\item[] $\mathbf{a} = n \times 1$ vector of additive genetic values,

\item[] $\mathbf{e} = n \times 1$ vector of environmental effect,

\item[] $\mathbf{b} = p \times 1$ vector of SNP effects assumed $\sim$$\mathrm{N}(0,\sigma_{b}^{2})$,

\item[] $\mathbf{W} = \mathrm{an}$ incidence matrix allocating SNP effects to individuals.
\end{itemize}
The commonly used ``animal model'' for estimating genetic value in
livestock and natural populations is also $\mathbf{y} = \mathbf{a} +
\mathbf{e}$ as above, but with $V(\mathbf{a}) =
\mathbf{A}\sigma_{a}^{2}$, where $\mathbf{A}$ is the numerator
relationship matrix defined by the relationships between individuals
known from their pedigrees. Thus, our model is the same as the normal
animal model but with the relationships between individuals estimated
from the markers $(\mathbf{WW}')$ rather than from the pedigree \cite{43}. The
$\mathbf{A}$ matrix assumes that an individual inherits exactly
$\tfrac{1}{4}$ of its genes from each grandparent. Although this is
correct on average, individuals deviate from this expectation and
$\mathbf{WW}'$ tracks these deviations from expectation. Thus, the
prediction of phenotype described above uses the average relationship
between individuals and deviations from this relationship specific to a
part of the genome. For example, Visscher et al. (2006) \cite{44} showed that
these deviations from the average relationship among full-sibs could be
used to estimate heritability and Hayes et al. (2009) \cite{45} showed that
the phenotype of a new sibling could be predicted accurately if the
reference sample contained a large number of full-sibs.

A GWAS among members of a family (e.g., a family of full-sibs) would
normally be described as a linkage analysis. In such an analysis markers
some distance from a QTL would show an association with the trait
because there has only been one generation of recombination between the
parents (or common ancestors) and the full-sibs. Consequently, a marker
allele and a QTL allele on the same chromosome will be inherited
together much of the time. Most GWAS in humans use individuals who have
no known relationship and are presumably only distantly related.
However, distant common ancestors still exist and markers closely linked
to the QTL will still be inherited together with the QTL in modern
descendants of these common ancestors. GWAS in livestock typically use
related animals and so the common ancestors are more recent, the size of
chromosome segments inherited from these common ancestors is greater and
so markers a greater distance from the QTL will show an association with
the trait. In general, QTL-marker disequilibrium is larger in
populations with a smaller effective population size.

\subsection{Application of Prediction of Phenotype from SNPs} \label{sec:3.4}

Meuwissen et al. (2001) \cite{30} presented methods to predict genetic value
for a complex trait based on genome wide markers. Using simulated data,
they reported correlations between predicted genetic value and true
genetic value as high as 0.85. In practice, such high values have not
been obtained, although Van Raden et al. (2008) \cite{46} reported a
correlation of 0.7 for milk yield in dairy cattle and Lee et al. (2008)
\cite{41} a correlation of 0.4--0.8 for three quantitative traits in mice. Two
features of the simulation in Meuwissen et al. (2001) \cite{30} favored
highly accurate prediction of genetic value. First, the population
simulated had an effective population size ($N_{e}$) of 100 and hence a
high level of LD between the markers and the QTL. Second, the authors
based their prediction on haplotypes of two multi-allelic markers. These
haplotypes should explain more of the variation at QTL than single SNP
markers which are bi-allelic. The simulation study also confirmed that
the accuracy was highest when the distribution of effects used in the
prediction of genetic value matched the true distribution (that was used
to simulate the data).

\begin{figure*}[t]

\includegraphics{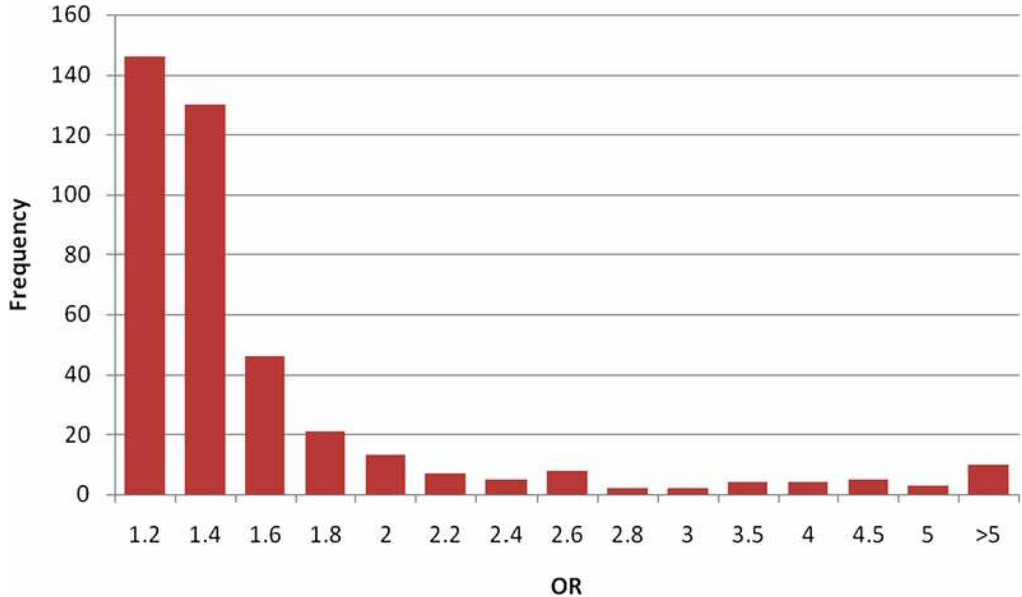}

  \caption{Distribution of validated effect sizes identified in
case-control GWAS \protect\cite{56} for a range of common diseases.}\label{fig3}
\end{figure*}

When the results of GWAS are published they usually focus on a few
highly significant SNPs. But GWAS also contain information about the
overall genetic basis for complex traits, for example, the number of
genes affecting a trait, the number of polymorphisms at these genes,
their allele frequencies and their effects on the trait. This
information is of interest in its own right and is also useful in
setting prior distributions such as $p(b)$ used in predicting phenotype.
Figure 3 presents the distribution of SNP effects for a range of
diseases published to date from GWAS. These estimates are from the
validation experiment so that they minimize the bias described above for
significant SNPs in the discovery experiment. Typically, SNPs increase
the risk of disease by 1.1--1.3. The number of independent SNPs of this
effect size needed to explain all the observed genetic variance depends
on the allele frequencies of the SNPs and the genetic variance of the
trait. Assuming the SNPs that have been discovered and reported are
typical in effect size and allele frequencies, 100--1000 SNPs would be
needed to explain the genetic variance of the diseases in Table 1 \cite{25}.
In fact, the SNPs discovered are likely to have larger than average
effect sizes and so the total number of genes needed to explain the
observed genetic variance is probably very large. Similar conclusions
can be reached from the published GWAS on human height. The effect sizes
are small (0.1--0.3 cm per allele) and so 100s of such SNPs are needed to
explain the genetic variance of height \cite{47}.

There have been no published reports of the accuracy of predicting
complex traits from GWAS in humans. However, 41 SNPs associated with
human height have been reported from 3 large GWAS. Collectively, these
explain only $\sim $6\% of the variance, giving a correlation between
phenotype of predicted genotype of $\sqrt{(0.06)} = 0.23$. This is
disappointing and typical of other traits, leading to the question
``Where are the missing genes?'' \cite{48}. Even when we have attempted to
use all available SNPs, we do not observe high correlations (Yang,
Goddard, Visscher and others, unpublished).

Goddard (2009) \cite{43} and Hayes et al. (2009) \cite{45} developed analytic
methods to calculate the accuracy of prediction of genetic value from
markers. A small $N_{e}$, a small number of QTL affecting the trait, a
high heritability, a large number of markers and a large number of
individuals in the reference sample lead to a high correlation between
predicted and true genetic value.

Which of the statistical methods gives the most accurate prediction
depends on which distribution of $b$ corresponds most closely to the
real world. In the analysis of data on milk production traits in cattle,
Van Raden et al. (2008) \cite{46} found that other methods gave only a small
improvement over BLUP, implying that many SNPs, each with a small
effect, give the best prediction. Lee et al. (2008) \cite{41}, using data on
mouse coat color and other quantitative traits and a method similar to
Bayes B, found that only a small proportion of markers were needed, as
might be expected given the small number of known genes affecting coat
color. More markers were needed for two other quantitative traits.

$N_{e}$ in dairy cattle has been about 100 for the last 6--40 generations
but was 1000--2000 prior to that \cite{49}. On the other hand, $N_{e}$ in
humans went through a bottleneck of about 3000, approximately 500
generations ago, and has since expanded enormously \cite{50}. The recent
small $N_{e}$ in cattle has generated some long range LD which increases
the accuracy of prediction. The equivalent model described above
suggests another way to describe this situation: in cattle there are
many individuals that have inherited the same chromosome segment from a
common ancestor and the markers track this relationship. In most human
populations there are few of these close relationships and so the
markers cannot so easily track the inheritance of identical chromosome
segments. The mouse data came from the heterogeneous mouse line \cite{51}
derived from crossing inbred strains and consequently has long range LD,
making it possible for markers to predict genetic value by tracing large
chromosome segments.

The milk yield data analyzed by VanRaden et al. (2008) \cite{46} was the
average milk yield of the daughters of each bull and, consequently, the
heritability of these ``phenotypes'' is high, approximately 0.8. The
heritability of human height is also about 0.8, so this does not explain
the lower accuracy of predicting height \cite{52}. However, a number of
diseases have lower heritability and this partly explains the difficulty
in predicting them. VanRaden et al. (2008) \cite{46} reported that
increasing the number of individuals and increasing the number of
markers both increased the accuracy of predicting genetic value, as
would be expected from the theory described here and elsewhere.

\section{Future Developments} \label{sec:4}

The theory in Goddard (2009) \cite{53} and Hayes et al. (2009) \cite{43} is
supported by the limited data on the accuracy with which genetic value
can be predicted from SNPs. However, the proportion of genetic variance
explained by SNPs for human height is still lower than expected, begging
the question ``Where are the missing genes''? Future research must try
to answer this question. The explanation that there is substantial
epistasis, or that \textit{de novo} mutants (including \textit{de novo}
copy number variants) or epigenetic effects are important, is
unsatisfactory. The heritability of human height ($\sim $0.8) is the
narrow sense heritability, that is, it is the proportion of phenotypic
variance that is due to the additive effect of genes. Epistasis does not
contribute to the narrow sense heritability, \textit{de novo} mutations
are by definition not inherited and few inherited epigenetic changes are
known. In any case, an inherited stable epimutation, for example, a
mutation that changes the methylation status at a locus and affects the
phenotype, would in practice behave just like a mutation that changes a
nucleotide, in terms of the resemblance between relatives and the
SNP-phenotype correlation in gene mapping experiments.

There appear to be three possible explanations. First, the genetic
markers (i.e., SNPs) may not be tracking the QTL. That is, the SNPs may
not be in high enough LD with the QTL. Since we have not identified most
of the QTL, we cannot answer this question directly. If we assume that
the QTL are similar to SNPs in their properties, we can use the LD
between SNPs as a guide to the LD between SNPs and QTL. The SNPs on
commercial ``SNP chips'' are considered to represent about 68--92\% of
the known common genetic \mbox{variation} when compared to variation in samples
representing 120 Caucasian chromosomes genotyped in the human HapMap
project \cite{54}. However, near complete sequencing of 76 genes on the same
\mbox{subjects} \cite{55} has identified more common variants, suggesting that only
57--79\% of common variation is represented by the current generation of
SNP chips. Thus, the coverage is good but not perfect. However, QTL may
have different properties to SNPs. For example, they may be under
stronger selection, and therefore be younger polymorphisms with lower
minor allele frequency. This would decrease LD with SNP markers.
Alternatively, QTL may often be deletions or duplications of DNA which
interfere with the ability to assay SNPs near enough to be in LD with
the deletion or duplication. Some copy number variants (DNA sequences
which vary between individuals in the number of copies they carry on a
chromosome, e.g., insertion/deletion variants) can also be typed using
the latest generation of SNP chips.

Second, the variance explained by individual QTL may be so small that
experiments with 10,000s of individuals are not powerful enough. The
finding in dairy cattle, that a prediction method designed for the
situation where all SNPs have small effects performs well, gives some
support to this explanation. Even if every one of the $3 \times  10^{9}$ bases
of DNA had a tiny effect on a trait, LD among these QTL would create
``\mbox{super-loci,}'' each consisting of a chromosome segment inherited as a
block and it would only be necessary to estimate the combined effects of
the QTL on this segment. As pointed out by Goddard (2009) \cite{43}, the size
of these segments depends on the $N_{e}$ of the population and the
recombination rate. The genome can be divided into approximately
$4N_{e}L$ segments or effective QTL, where $L$ is the length of the
genome in recombination units or Morgans. Thus, a population with a
large $N_{e}$ can have a large number of effective QTL but a population
with small $N_{e}$ cannot.

Both of these first two explanations of the low accuracy with which
genetic value can be predicted from SNPs might apply to humans. The
third explanation is that there are some phenomena explaining
inheritance of which we are totally unaware.

Assuming that the first two explanations are enough, we should be able
to explain a larger proportion of the genetic variance by increasing the
number of individuals in the reference sample and by increasing the
density of markers. In fact, complete sequencing of the genome is likely
to replace or complement genotyping of known polymorphism in the
foreseeable future. This should allow causal polymorphisms to be used in
prediction instead of linked markers. However, it will also increase the
number of variable sites whose effects must be estimated by an order of
magnitude or more. Despite this increase in data quantity and quality,
the methods of predicting genetic value and of estimating the effect of
polymorphisms, discussed in this paper, will still be relevant. We
expect that as the number of polymorphisms increases and includes the
causal variants, methods of prediction that assume many markers have no
effect on the trait will perform better than methods that assume all
markers have some effect. This is because we expect markers with no
direct effect to cease to be useful predictors when the causal
polymorphism, to which they are linked, is included in the model.
However, whether the data will be sufficiently powerful to distinguish
causal variants from markers in LD with them remains to be seen.

\section*{Acknowledgments}

This work was supported by the Australian National and Medical Research
Council Grants 389892, 442915, 339450, 443011 and 496688.

\end{document}